\documentclass[a4paper,fleqn]{cas-dc}

\usepackage[numbers, sort&compress]{natbib}
\usepackage{diagbox}
\usepackage{color}
\usepackage{listings}
\usepackage{lipsum}
\usepackage{caption}

\def\tsc#1{\csdef{#1}{\textsc{\lowercase{#1}}\xspace}}
\tsc{WGM}
\tsc{QE}
\begin{document}
\let\WriteBookmarks\relax
\def\floatpagepagefraction{1}
\def\textpagefraction{.001}

\shorttitle{}    

% \shortauthors{Liu, Zhang}

\title [mode = title]{Physics-Informed Neural Networks for the Quantum Droplets in Binary Bose-Einstein Condensates}  


\author[1,2]{Dongshuai Liu}
% \ead{dongs_liu@fynu.edu.cn}
\author[3]{Boris A. Malomed}
\author[4]{Wen Zhang \corref{cor}}% \fnmark[1]
\ead{202406034@fynu.edu.cn}

% \fntext[1]{These authors contributed equally to this work.}
\cortext[cor]{Corresponding author}

\affiliation[1]{organization={School of Physics and Electronic Engineering, Fuyang Normal University, Fuyang 236037, China}}
\affiliation[2]{organization={Key Laboratory of Functional Materials and Devices for Informatics of Anhui Educational Institutions, Fuyang Normal University, Fuyang, Anhui 236037, China}}
\affiliation[3]{organization={Instituto de Alta Investigaci\'{o}n, Universidad de Tarapac\'{a}, Casilla 7D, Arica, Chile}}
\affiliation[4]{organization={School of Mathematics and Statistics, Fuyang Normal University, Fuyang 236037, China}}

% \affiliation[5]{organization={Department of Physical Electronics, School of Electrical Engineering, Faculty of Engineering, Tel Aviv University, Tel Aviv 69978, Israel}}

\begin{abstract}   
Physics-Informed Neural Networks (PINNs), which integrate deep learning with physical prior knowledge, have proven to be a powerful tool for studying the dynamics of high-dimensional nonlinear systems. The present work utilizes PINNs to analyze the existence and evolution of quantum droplets (QDs) in a binary Bose-Einstein condensate (BEC), revealing the ability of this technique to accurately predict structural features of the QDs, their multipeak profiles, and dynamical behavior. The stable evolution of multipole QDs is thus demonstrated. Comparing different network architectures, including the training time, loss values, and $\mathbb{L_{2}}$ error, PINNs accurately predict specific dynamical characteristics of QDs. Furthermore, the PINN robustness is evaluated by the application of PINN to parameter-discovery tasks, considering both clean training data and data contaminated by $1\%$ random noise. The results highlight the efficiency of PINNs in modeling complex quantum systems and extracting reliable parameters under the noisy conditions.
\end{abstract}

% Keywords
\maketitle
% \renewcommand{\thefootnote}{\fnsymbol{footnote}}
% % \footnotetext[1]{Sabbatical address}
% \footnotetext[2]{Corresponding author}
% Main text
\section{Introduction}

\label{}

Quantum droplets (QDs) are localized states filled by ultradilute but nearly incompressible superfluids, that arise from the interplay of Hamiltonian terms of two types: the mean-field (MF) energy and the beyond-MF quantum correction, alias the Lee-Huang-Yang (LHY) effect \cite{lee_eigenvalues_1957}, in a weakly interacting binary Bose-Einstein condensate (BEC). The balance between the cubic MF intracomponent repulsion and intercomponent attraction between atoms can be precisely controlled by means of the Feshbach-resonance technique \cite{thalhammer_double_2008, roy_test_2013, wang_double_2015}. The LHY correction gives rise to the additional quartic self-repulsion, which prevents the onset of the collapse and stabilizes two-component QDs \cite{petrov_quantum_2015, dongThreedimensionalVortexMultipole2024a}. The superfluid in QDs features the density up to eight orders of magnitude lower than liquid helium \cite{schmitt_self-bound_2016, kartashov_bloch_2016, cabrera_quantum_2018, de_rosi_thermal_2021}. QDs are not only observed at the macroscopic scale in atomic BEC \cite{semeghiniSelfBoundQuantumDroplets2018b,derricoObservationQuantumDroplets2019c}, but also appear in nuclei \cite{noauthor_nuclear_nodate} and superfluid helium \cite{ancilotto_spinning_2018}.

If the effective dimension is reduced from 3D to 1D, the LHY term in the respective Gross-Pitaevskii equation (GPE) becomes quadratic, with the opposite sign, corresponding to self-attraction, while the residual cubic MF interaction is then considered in the regime of self-repulsion \cite%
{petrov_ultradilute_2016, astrakharchik_dynamics_2018}. In 2D the LHY correction gives rise to the logarithmic factor multiplying the cubic term \cite{petrov_ultradilute_2016, dong_rotating_2021}.

Unlike the famous 1D nonlinear Schr\"{o}dinger equation \cite{shabat1972exact}, the LHY-amended GPEs are not integrable. Therefore, numerical studies based on these equations \cite{li_two-dimensional_2018, kartashov_metastability_2019, skov_observation_2021, liuRotatingDipoleQuadrupole2024b} and the diffusion Monte-Carlo technique \cite{parisi_liquid_2019, cikojevic_universality_2019, cikojevic2020finite} have been performed for QDs in different spatial dimensions. While traditional numerical methods, such as the finite-difference algorithm, Newton's iterations, and spectral schemes, have been used to study the dynamics of solitons, these conventional techniques face significant challenges when solving nonlinear equations without well-defined initial and boundary conditions. Additionally, the traditional numerical techniques often struggle with complex inverse problems, particularly those involving nonlinearity, non-convexity, or high dimensionality.

With the rapid development of computational resources and artificial intelligence, data analysis and deep-learning techniques have become powerful tools for handling sophisticated nonlinear equations \cite{luDeepXDEDeepLearning2021}. These methods, based on neural networks, efficiently address nonlinear equations through the data feature extraction and function approximation. The emergence of Physics-Informed Neural Networks (PINNs), which incorporate prior physical knowledge, marks a significant breakthrough in this field. The PINN scheme fundamentally differs from other neural networks, such as convolutional ones \cite{huangMachineLearningAssistedQuantumControl2022}, which are purely data-driven, do not account for physical constraints, and are typically applied to image processing and recognition tasks. On the other hand, PINNs integrate physically relevant equations directly into the neural network architecture, ensuring that the solutions learned by the network adhere to physical laws \cite{raissi_physics-informed_2019}. This approach has been widely applied across various fields of research: computational fluid dynamics \cite{zhang_cpinns_2023, caiPhysicsinformedNeuralNetworks2021, zhang_robust_2023}, nonlinear dynamical systems \cite{jiang_physics-informed_2022, wu_efficient_2024, luPhysicsinformedNeuralNetworks2024, liuPhysicsinformedNeuralNetwork2024, song_two-stage_2024}, and reaction-diffusion systems \cite{sun_physics-informed_2023, zhang_discovering_2024}, among others. With continuous optimization of PINNs architectures and training methods, this approach has significantly improved the accuracy and efficiency in solving direct and inverse problems based on nonlinear partial differential equations.

The application of PINNs to study QDs in BECs has gained significant interest. PINNs, by incorporating physical laws directly into the structure of neural networks, have been effectively utilized to model dynamics of QDs, including their formation, stability, and evolution \cite{songDatadrivenTwodimensionalStationary2024, puComplexDynamicsOnedimensional2023}. The method has been shown to provide accurate predictions of QDs profiles and dynamics, even in the presence of noise or incomplete data. Furthermore, the use of PINNs allows for the discovery of key system parameters from both clean and perturbed data, offering new insights into the physics of BECs and the broader field of nonlinear quantum fluids \cite{jiangPredictionSymmetricAsymmetric2024a}. Through continuous refinement of the PINN architecture and training techniques, these techniques
significantly improve understanding of the formation and stability of QDs in ultra-cold atomic gases.

Although significant progress has been made in the studies of QDs, the existence and dynamics of QDs in the free space is far from complete. The current work primarily focuses on trapped configurations under the action of external potentials, highlighting a gap in understanding the behavior of QDs in the absence of external constraints. In particular, a natural question arises: can the PINN method effectively predict the free-space evolution of multi-peak QDs in binary BECs.

In the present work, the PINN method is employed to investigate the existence, evolution, and robustness of fundamental and multipole QDs in binary BECs. Notably, multipole QDs exhibit multi-peak structures with narrow profiles, which are found to be stable. By comparing different network architectures, including training time, loss values, and $\mathbb{L_{2}}$ error, we have found that setups with few network layers can still accurately predict the evolution of QDs. Additionally, leveraging known stationary solutions for QDs, a data-driven parameter discovery of the GPE was conducted. Even with $1\%$ random noise added to the stationary solutions, the PINN method successfully predicts unknown system's parameters. These results highlight the superior performance of the PINN method in solving forward problems and predicting the QD dynamics.

%%%%%%%%%%%%%%%%%%%%%%%%%%%%%%%%%%%%%%%%%%%%%%%%%%%%%%%%%%%%%%%%%

\section{The physical model and methodology}

\label{}

\subsection{Theoretical model}

We consider QDs in the binary condensates with two components characterized by coupling constants ${g}_{\upuparrows (\downdownarrows )},{g}_{{\uparrow }{\downarrow }}$, which represent the intra-component repulsion and the inter-component attraction, respectively. In the vicinity of the MF collapse instability point, the energy density of the system is expressed as \cite{petrov_ultradilute_2016}:

\begin{gather}
\epsilon=\frac{1}{2}\left( \sqrt{{g}_{\upuparrows }}n_{\uparrow }-\sqrt{{g}%
_{\downdownarrows }}n_{\downarrow }\right) ^{2}  \notag \\
+\frac{\sqrt{{g}_{\upuparrows }+{g}_{\downdownarrows }}\delta g(\sqrt{{g} _{\downdownarrows }}n_{\uparrow }+\sqrt{{g}_{\upuparrows }}n_{\downarrow})^{2}}{({g}_{\upuparrows }+{g}_{\downdownarrows })^{2}}  \notag \\
-\frac{2}{3\pi }({g}_{\upuparrows }n_{\uparrow }+{g}_{\downdownarrows
}n_{\downarrow })^{{3}/{2}},  \label{eq:refname1}
\end{gather}
where $n_{\uparrow }$ and $n_{\downarrow }$ are the densities of the two components, and $\delta {g}={g}_{{\uparrow }{\downarrow }}+\sqrt{{g} _{\upuparrows }{g}_{\downdownarrows }}>0$. The minimization of the MF contribution of the first term of expression \eqref{eq:refname1} forces the components of the binary mixture to obey the radio $n_{\uparrow }/n_{\downarrow }=\sqrt{{g}_{\upuparrows }/{g}_{\downdownarrows }}$, which is also satisfied in inhomogeneous configurations with smooth variations of the total density \cite{tylutki_collective_2020}. Thus, the two-component binary BEC is effectively reduced to the single-component system. We assume that the coupling constants of the two components are equal, i.e., ${g} _{\upuparrows }={g}_{\downdownarrows }\equiv {g}$, and the numbers of atoms in the components are equal too. Accordingly, the resulting 1D GPE for the with MF wave function $\Psi $, with the MF cubic self-repulsion and quadratic LHY attraction is reduced to the known equation \cite{astrakharchik_dynamics_2018}, written in the scaled form:
\begin{equation}
{i}\frac{\partial \Psi }{\partial {T}}=-\frac{1}{2}\frac{\partial ^{2}\Psi }{%
\partial {X}^{2}}-\frac{\sqrt{2}}{\pi }{g}^{3/2}|\Psi |\Psi +\delta {g}|\Psi
|^{2}\Psi ,  \label{eq:refname2}
\end{equation}%
with the effective coupling constant $g\equiv \sqrt{{g}_{\upuparrows }{g} _{\downdownarrows }}>0$.

We further simplify Eq. \eqref{eq:refname2} by introducing the following units of length $x_{0}$, time $t_{0}$, and normalization factor $\psi _{0}$ for the MF wave function $\Psi $:
\begin{equation}
\begin{split}
& \psi _{0}=\frac{\sqrt{2}{g}^{3/2}}{\pi \delta {g}},\Psi =\psi _{0}\psi ; \\
& {x}_{0}=\frac{\sqrt{\delta {g}}\pi }{\sqrt{2}{g}^{3/2}},{X}={x}_{0}{x}; \\
& {t}_{0}=\frac{\delta {g}\pi ^{2}}{2{g}^{3}},{T}={t}_{0}{t}.
\end{split}
\label{eq:refname3}
\end{equation}%
The so normalized GPE takes the form of
\begin{equation}
{i}\frac{\partial \psi }{\partial {t}}=-\frac{1}{2}\frac{\partial ^{2}\psi }{%
\partial {x}^{2}}-|\psi |\psi +|\psi |^{2}\psi ,  \label{eq:refname4}
\end{equation}

Stationary solutions for self-bound QDs are looked for by substituting the usual ansatz,
\begin{equation}
\psi =\phi (x)\exp (-i\mu t),  \label{eq:refname5}
\end{equation}%
where $\mu $ is chemical potential and $\phi (x)$ determines the QD profile, which vanishes at $x\rightarrow \pm \infty $. The total norm $N$ of the solution, which is proportional to the number of atoms in the condensate, is ${N}=\int_{-\infty }^{+\infty }|\phi |^{2}dx$, and its energy is
\begin{equation}
E=\int_{-\infty }^{+\infty }\left[ \frac{1}{2}\left\vert \frac{d\phi }{dx}%
\right\vert ^{2}-\frac{2}{3}\lvert \phi |^{3}+\frac{1}{2}|\phi |^{4}\right]
dx.
\end{equation}

The formation of stable QDs is provided by the balance between the MF repulsive cubic nonlinearity and the attractive LHY correction \cite{kartashovEnhancedMobilityQuantum2024}. Further, multi-droplet structures manifest this equilibrium in the spatially extended form: while each constituent QD remains stable in a manner similar to the stability of an isolated one, the overall stability of the QD array is sustained by inter-QD interactions as described above. Our objective is to use the PINN technique for predicting the existence and dynamics of QDs in the binary BEC. To this end, we employ stable multipole QDs as the prior knowledge for training the neural network.

\subsection{PINN methodology}

According to Eq.~\ref{eq:refname4}, we first define the residual neural network $F _{1}(t,x)$ as
\begin{equation}
F_{1}(x,t)\equiv i\frac{\partial \phi }{\partial {t}}+\frac{1}{2}\frac{%
\partial ^{2}\phi }{\partial x^{2}}+|\phi |\phi -|\phi |^{2}\phi ,
\label{F1}
\end{equation}%
where $\widehat{\phi }(x,t)=\widehat{u}(x,t)+i\widehat{v}(x,t)$ is the QD complex solution that should be predicted. Thus, $F_{1}(x,t)$ splits in the real and imaginary components:
\begin{equation}
\begin{split}
& F_{\mathrm{re}}=-\widehat{v}_{z}+\frac{1}{2}\frac{\partial ^{2}\widehat{u}%
}{\partial x^{2}}+\sqrt{\widehat{u}^{2}+\widehat{v}^{2}}\widehat{u}-(%
\widehat{u}^{2}+\widehat{v}^{2})\widehat{u}, \\
& F_{\mathrm{im}}=-\widehat{u}_{z}+\frac{1}{2}\frac{\partial ^{2}\widehat{v}%
}{\partial x^{2}}+\sqrt{\widehat{u}^{2}+\widehat{v}^{2}}\widehat{v}-(%
\widehat{u}^{2}+\widehat{v}^{2})\widehat{v}.
\end{split}
\label{Freim}
\end{equation}%
The predicted solution is incorporated into the PINN via Eq.~\eqref{Freim}, ensuring that a proper physical interpretation of $\widehat{\phi }(x,t)$ is imposed by physical constraints to avoid overfitting. 

Overfitting occurs when a model learns not only the underlying physical relationships of the data, but also fits the noise and random fluctuations present within it. A model that experiences overfitting may produce a solution $\widehat{\phi}(x,t)$ that perfectly matches the training data, yet potentially violates the physical law $\mathcal{F}$ described by Eq.~\eqref{Freim}. This causes the model to degenerate into a mere ``look-up tool" for the data, thereby losing its generative capability. In the framework of PINNs, by incorporating the physical equation \eqref{Freim} as a constraint, the network is penalized for generating solutions that violate the physical laws. This feature forces the network to seek a solution $\widehat{\phi}(x,t)$ satisfying both the sparse data points and the physical constraints, thus resulting in a more robust and reliable model, which demonstrates a superior generative performance.

To optimize the PINN hyperparameters, including weights and biases, the loss function is defined using the mean-squared error (MSE):
\begin{equation}
{Loss}=\mathrm{MSE}_{\mathrm{IC}}+\mathrm{MSE}_{\mathrm{BC}}+\mathrm{MSE}_{%
\mathrm{F}},  \label{Loss}
\end{equation}%
where $\mathrm{MSE}_{\mathrm{IC}}$ quantifies the difference between the initial conditions (ICs) $\phi (t=0,x_{0}^{i})$ and the predicted results $\widehat{\phi }(t=0,x_{0}^{i})$ in the framework of PINN. Further, $\mathrm{MSE}_{\mathrm{BC}}$ in Eq. \eqref{Loss} accounts for the periodic boundary conditions (BCs), while $\mathrm{MSE}_{\mathrm{F}}$ captures the residual error from randomly selected points in the space-time domain. These terms are computed as:
\begin{equation}
\begin{split}
& \mathrm{MSE}_{\mathrm{IC}}=\frac{1}{N_{\mathrm{IC}}}\sum_{i=1}^{N_{\mathrm{%
IC}}}|\widehat{\phi }(x_{0}^{i},0)-\phi (x_{0}^{i},0)|^{2}, \\
& \mathrm{MSE}_{\mathrm{BC}}=\frac{1}{N_{\mathrm{BC}}}\sum_{i=1}^{N_{\mathrm{%
BC}}}[|\widehat{\phi }(x_{1},t_{BC}^{i})-\phi (x_{2},t_{BC}^{i})|^{2} \\
& \qquad +|\widehat{\phi }(x_{BC}^{i},t_{1})-\phi (x_{BC}^{i},t_{2})|^{2}],
\\
& \mathrm{MSE}_{\mathrm{F}}=\frac{1}{N_{\mathrm{F}}}%
\sum_{i=1}^{N_{F}}|F_{1}(x_{F}^{i},t_{F}^{i})|^{2},
\end{split}
\label{MSE}
\end{equation}%
where the observable measurements $\{{\widehat{\phi }(0,x_{0}^{i})}\}_{i=1}^{N_{\mathrm{IC}}}$ represent the sampled data at $t=0$ provided by the ICs. Here, $N_{\mathrm{IC}}$ and $N_{\mathrm{BC}}$ denote the number of training points on the ICs and BCs, respectively, while $N_{\mathrm{F}}$ refers to the number of collocation points. The terms $\{\widehat{\phi }(t_{\mathrm{BC}}^{i},x)\}_{i=1}^{N_{\mathrm{BC}}}$ and $\{\widehat{\phi }(t_{\mathrm{BC}}^{i}, x_{\mathrm{BC}}^{i})\}_{i=1}^{N_{\mathrm{BC}}}$ correspond to the latent QDs solution associated with the boundary training data, and $\{t_{F}^{i},x_{F}^{i}\}_{i=1}^{N_{F}}$ are the space-time points for training the residual $F_{1}(t,x)$. In particular, to generate the training data, the Newton-conjugate-gradient method \cite{yang2010nonlinear} is employed to solve Eq. \eqref{eq:refname4} in the numerical form. After obtaining the stationary solution at $t=0$, we shift our focus to the initial-boundary value problem for QDs. Additionally, the split-step method incorporating absorbing BCs is applied to generate a high-resolution training dataset for predicting the evolution of QDs by the PINN. The PINN sampling points are selected using the space-filling Latin Hypercube Sampling (LHS) strategy \cite{stein_large_1987}. All training was performed on a desktop computer equipped with an Intel Xeon CPU E5-2690 v4 @2.60GHz and an NVIDIA GeForce RTX 2080 Ti.

The PINNs technique minimizes the loss function by optimizing the slope of the activation function, as well as the weights and biases, to generate high-resolution, predictable multipole QDs $\widehat{\phi }(t,x)$ that closely approximate genuine solutions ${\phi }(t,x)$. The Adam algorithm \cite{kingma_adam_2014}, known for its efficient first-order gradient optimization, is particularly well-suited for the application to large datasets and parameter-heavy problems. On the other hand, the limited-memory Broyden-Fletcher-Goldfarb-Shanno (L-BFGS) method \cite{liu_limited_1989} excels at handling multivariate optimization challenges.

In the framework of PINN, the shared parameters of the neural network are optimized by minimizing the MSE loss generated by the feedforward network $\widehat{\phi }(t,x)$ and the residual term $F_{1}(t,x)$. The optimization process combines the strengths of both the Adam and L-BFGS algorithms. Notably, the L-BFGS scheme enables faster network convergence with minimal fluctuations, significantly accelerating the PINN learning process. The L-BFGS optimization halts when the following iterative convergence criterion is satisfied:
\begin{equation}
\frac{|\mathrm{Loss}^{n}-\mathrm{Loss}^{n-1}|}{\max \{|\mathrm{Loss}^{n}|,|%
\mathrm{Loss}^{n-1},1|\}}<1.0\times \mathrm{np.finfo(float).eps},
\label{L-BFGS}
\end{equation}%
where $\mathrm{Loss}^{n}$ denotes the loss value at the $n$-th iteration, and $1.0\times \mathrm{np.finfo(float).eps}$ represents the Machine Epsilon. As parameter updates are independent of the gradient scaling process, the optimization begins with an efficient mini-batch Adam algorithm, followed by the application of a full-batch L-BFGS algorithm. This iterative process continues until the error of the loss function reaches machine precision, defined as per Eq. \eqref{L-BFGS}.

%%%%%%%%%%%%%%%%%%%%%%%%%%%%%%%%%%%%%%%%%%%%%%%%%%%%%%%%%%%%%%%%

\section{Data-driven multipole QDs}

\label{}

\begin{figure}
    \centering
    \includegraphics[width=\linewidth]{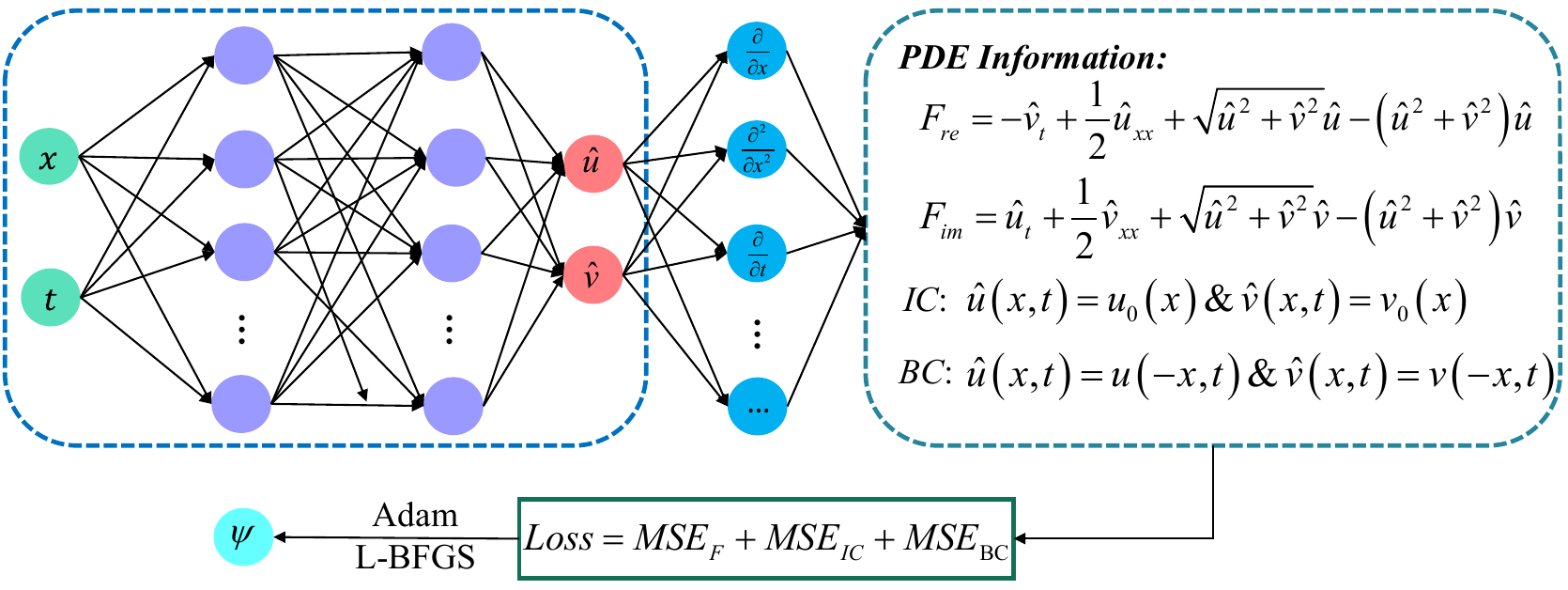}
    % Here is how to import EPS art
    \caption{The schematic PINN\ architecture for data-driven prediction of the evolution of the nonlinear system. The left panel includes the input, hidden, and output layers. The middle column represents the automatic differentiation technique in Tensorflow. The right panel shows the PINN dominated by the PDE construction, which corresponds to the PDE solution produced by the split-step method. The bottom corresponds to the loss function and optimization process (the Adam and L-BFGS optimizer).}
    \label{fig1}
\end{figure}

In this section, we apply the PINN method to study the evolution of multipole QDs in the binary BEC. The PINN with a multilayer deep neural network is constructed to approximate the droplet solution, as shown in Fig.~\ref{fig1}. The main framework of the PINN scheme consists of an input layer, several hidden layers with a certain number of neurons per layer, and an output layer. Output variables $\widehat{u}$ and $\widehat{v}$ must satisfy the ICs, BCs, and governing equations (the GPE). If the ICs are not satisfied, the tentative solution needs to be corrected instantaneously, so that the next moment can be predicted. Thus the loss function is made up of three parts that are involved in the PINN optimization process.

To satisfy the constraints of the initial-boundary conditions and governing equations, $N_{0}$ data points are randomly sampled within the ICs $\phi (t=0,x)$, and $N_{\mathrm{F}}$ auxiliary coordinates in the space-time domain are sampled via the LHS strategy with periodic BCs. The optimization algorithm fine-tunes the network parameters to minimize the loss function, thereby improving the network's performance. Through this process, we iteratively refine the network structure to reduce errors. After several iterations, the evolution of the multipole QDs in the binary BEC can be predicted with high accuracy.

First, we consider the evolution of monopole (fundamental) QDs. In the framework of GPE~\eqref{eq:refname4}, after obtaining the numerical solution for QDs, a split-step method with absorbing boundary conditions is used to generate high-resolution data. This dataset serves as the input for training the PINN to predict the evolution of the monopole QDs.

The evolution is revealed in the space-time domain $(t,x)\in \lbrack 0,2]\times \lbrack -15,+15]$. The spatial region is divided in $128$ equal segments along the $x-$axis, and $101$ nodes are fixed in the temporal interval. This means that the computed density $|\phi (t,x)|^{2}$ is represented by a matrix of size $101\times 128$. We use a five-hidden-layer neural network, each hidden layer consisting of $30$ neurons, to approximate the solution $\phi (t,x)$. The nonlinear activation function is chosen as $\tanh $ to represent all test scenarios. The output of the $k_\mathrm{th}$ layer in the neural network is given by $\phi ^{k} =\tanh(W ^k \phi ^{k-1} + b ^k)$, where $W ^k$ and $b ^k$ ($k=1,2 \cdots$) are the network's weight and bias term, respectively. The training set consists of $N_{0}=400$ randomly sampled points that form the initial BCs, accounting for approximately $3\%$ of the total data. According to the definition given by Eq.~\eqref{Freim}, the number of collocation points $F_{1}(t,x)$ for the PINN is set to be $N_{\mathrm{F}}=20000$. All sampling points are obtained across the entire space-time domain using the LHS strategy. For the Adam and multi-step L-BFGS optimization methods, the number of iterations is fixed to be $20000$ in both cases.

\begin{figure}
    \centering
    \includegraphics[width=\linewidth]{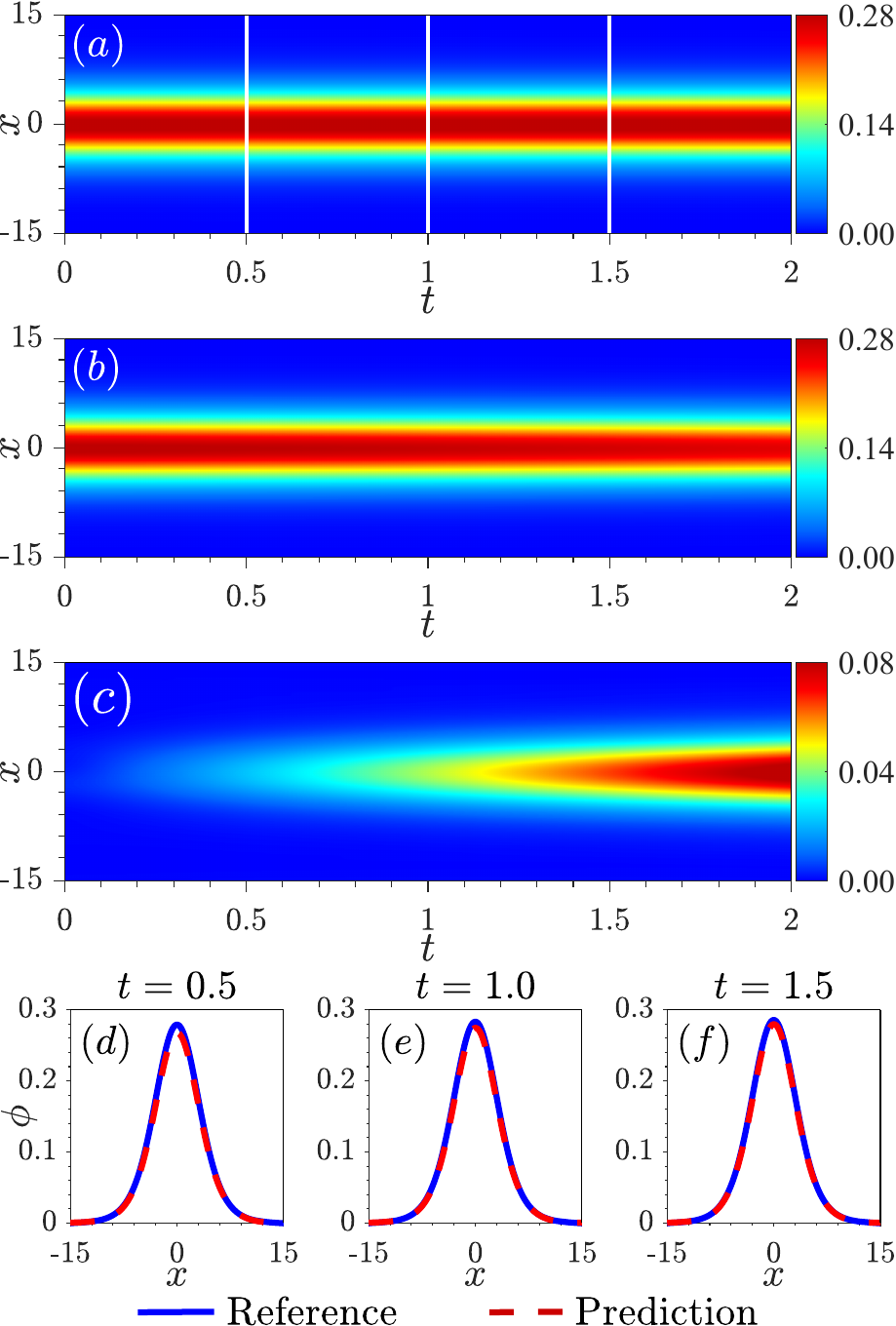}% Here is how to import EPS art
    \caption{The performance of the PINN for the density of the fundamental QDs. (a, b) The reference solution and the predicted output, respectively. (c) The squared error between the reference and predicted solution. (d)-(f) The comparisons between the profiles of the learned and reference fundamental QDs in three distinct snapshots, taken as $t=0.5,1.0$ and $1.5$, corresponding to the white vertical lines in (a).}
    \label{fig2}
\end{figure}

The evolution of the reference QDs and the predicted results for chemical potential $\mu =-0.15$ are shown in Figs.~\ref{fig2}(a) and (b), respectively. The monopole QDs exhibit the Gaussian-like structure and maintain stable evolution. From these figures, it is evident that the PINN method provides an accurate approximation for the monopole QDs in the binary BEC.

To further demonstrate the effectiveness of the PINN method, the absolute error between the exact and predicted solutions, $|\widehat{\phi }(t,x)-\phi (t,x)|$, is plotted in Fig.~\ref{fig2}(c), with residuals $\sim $ $10^{-2}$. Figures~\ref{fig2}(d)-(f) present the comparison of the predicted and numerically exact solutions at different times, $t=0.5$, $1$, and $1.5$. The total loss value, as defined by Eq.~\eqref{Loss}, is minimized to $0.9835\times 10^{-4}$, and the network achieves a relative $\mathbb{L}_{2}$ error of $0.4898\times 10^{-2}$ for $z=0$ at time $751.8777$ seconds. These results confirm that the predicted values match the numerically exact values of the density with high precision, underscoring the superiority of the PINN method in predicting the fundamental QDs solutions.

\begin{figure}
    \centering
    \includegraphics[width=\linewidth]{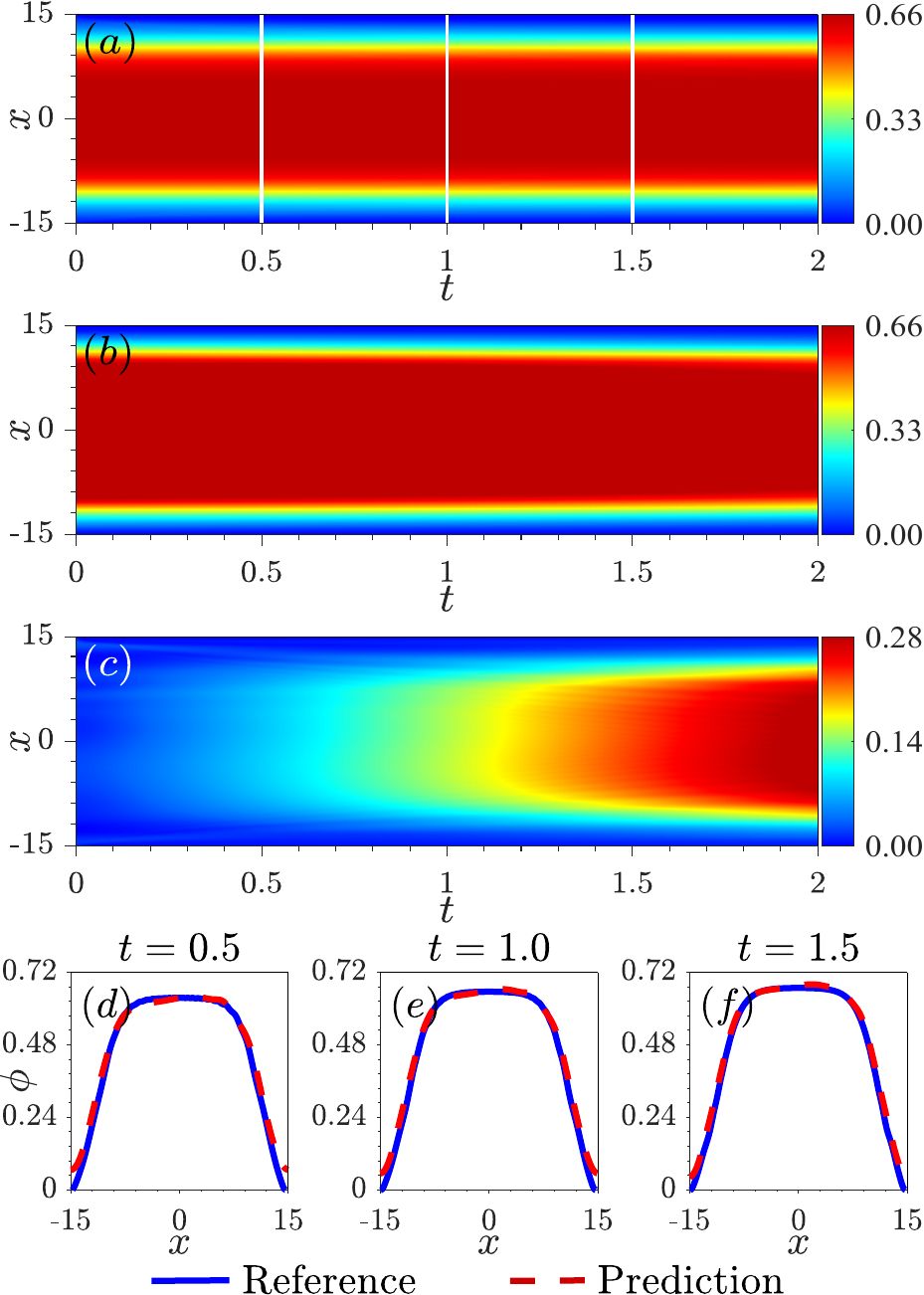}% Here is how to import EPS art
    \caption{The performance of the PINN for the density of the flat-top fundamental QDs. (a, b) The reference solution and the predicted output, respectively. (c) The squared error between the reference and predicted solution. (d)-(f) The comparisons between the profiles of the learned and reference fundamental QDs in three distinct snapshots, taken as $t=0.5,1.0$ and $1.5$, corresponding to the white vertical lines in (a).}
    \label{fig22}
\end{figure}

As the atom number increases, narrow Gaussian-shaped monopole droplets expand into broad flat-top ones. This transition occurs as well in one-dimensional QDs beyond the framework of the LHY approximation \cite{boudjemaaQuantumDropletsOnedimensional2023}. Thus, similar to the case of narrow monopole QDs, we have applied the PINN technique to predict flat-top QDs. The evolution of the reference QDs and the predicted ones, for a fixed chemical potential, $\mu =-0.22$, are shown in Figs.~\ref{fig22}(a) and (b), respectively. The absolute error between the exact and predicted solutions, $|\widehat{\phi }(t,x)-\phi (t,x)|$, is plotted in Fig. \ref{fig22}(c), with residuals $\sim $ $10^{-2}$. Figures~\ref{fig22}(d)-(f) present the comparison of the predicted and numerically exact solutions at different times, $t=0.5$, $1$, and $1.5$. The total loss value is minimized to $0.4361\times 10^{-4}$, and the network achieves a relative $\mathbb{L}_{2}$ error of $2.079\times 10^{-2}$ for $z=0$ at time $537.8562$ seconds. The results confirm that our PINN method accurately captures the characteristic flat-top density profile.

\begin{figure}
    \centering
    \includegraphics[width=\linewidth]{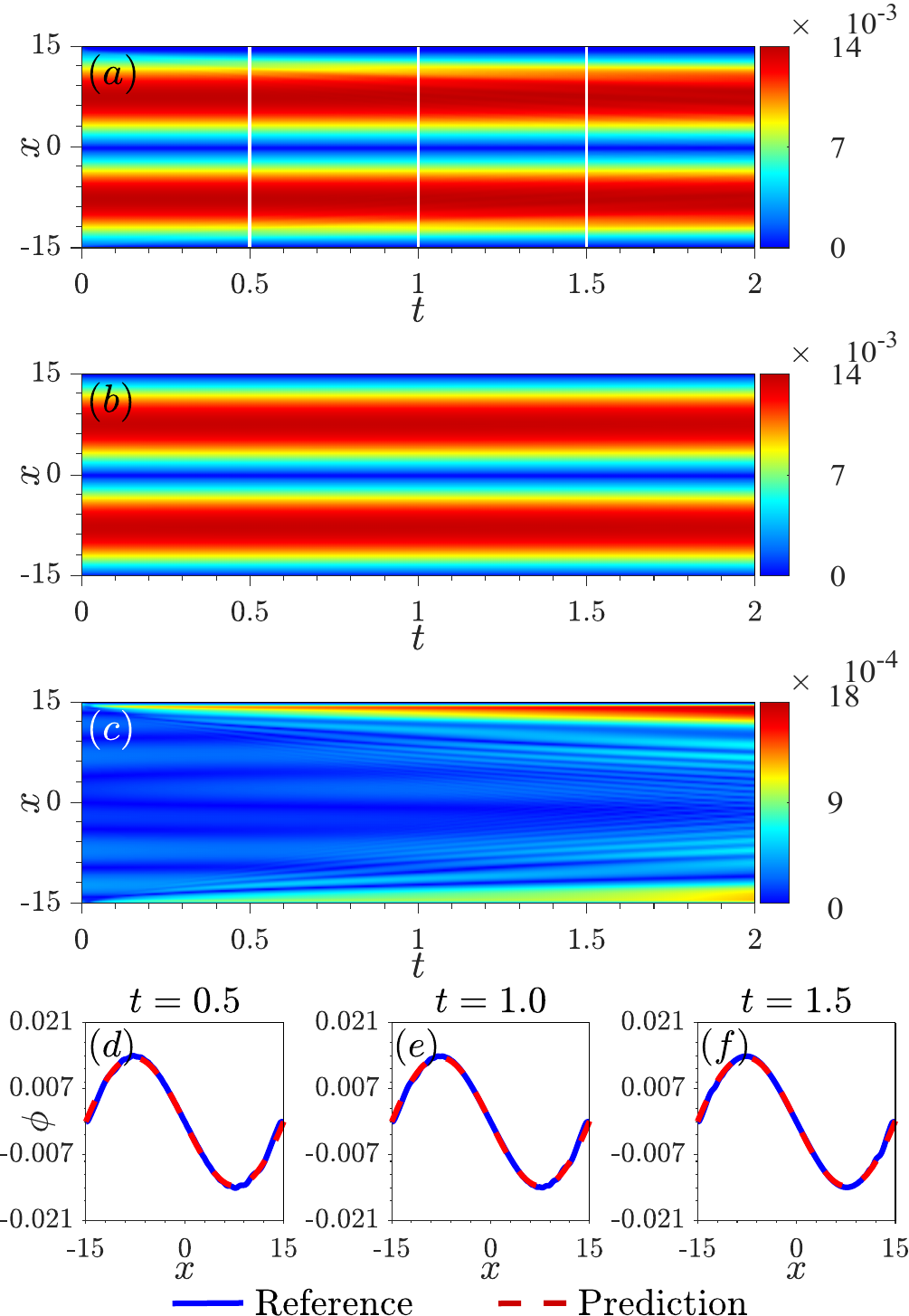}
    % Here is how to import EPS art
    \caption{The training results of dipole droplet. (a, b) The reference and learned solutions, respectively. (c) The squared error between the reference and learned solution. Panels (d)-(f) illustrate the comparisons between the reference solution and PINNs results at indicated value of time.}
    \label{fig3}
\end{figure}

The results produced by the PINN method for dipole QDs are shown in Fig.~\ref{fig3}. The sampling PINN dataset is produced by the Newton-conjugate and the split-step methods for $\mu =-0.15$. The entire spatiotemporal domain $(t,x)\in \lbrack 0,2]\times \lbrack -15,15]$ is divided in $101\times 128$ equal intervals along the $t$ and $x$ axes. The five-hidden-layer neural network is used, with each hidden layer consisting of $20$ neurons, to approximate the solution $\phi (t,x)$. For the initial-boundary conditions, $4.8\%$ of the dataset is chosen as sampling points. The LHS strategy is employed to generate $N_{\mathrm{F}}=20,000$ collocation points across the entire coordinate domain.

The numerically precise and predicted evolution of dipole QDs is shown, throughout the entire process, in Figs.~\ref{fig3}(a) and (b). Dipole QDs consist of two spatially separated monopole droplets with opposite phases, covering the entire window range. Compared to the monopole QDs, the density distribution in the dipole QDs is noticeably wider, with steeper edges. With the repulsion between their two constituents, dipole QDs are stable modes. The absolute error between the numerically exact and the predicted dipole-QD solutions is $\sim 10^{-4}$, as shown in Fig.~\ref{fig3}(c). Additionally, Figs.~\ref{fig3}(d)-(f) display the comparison between the numerical and PINN-predicted solutions at different times, $t=0.5$, $1$, and $1.5$, corroborating the accuracy of the prediction. After $20,000$ steps of the Adam learning and several L-BFGS optimizations, the network achieves a relative $\mathbb{L}_{2}$ error of $3.8068\times 10^{-2}$ at time $690.9437$ seconds, with the total loss $2.3105\times 10^{-7}$. Compared to the split-step method, the error throughout the entire propagation process confirms the reliability of the PINN method.

\begin{figure}
    \centering
    \includegraphics[width=\linewidth]{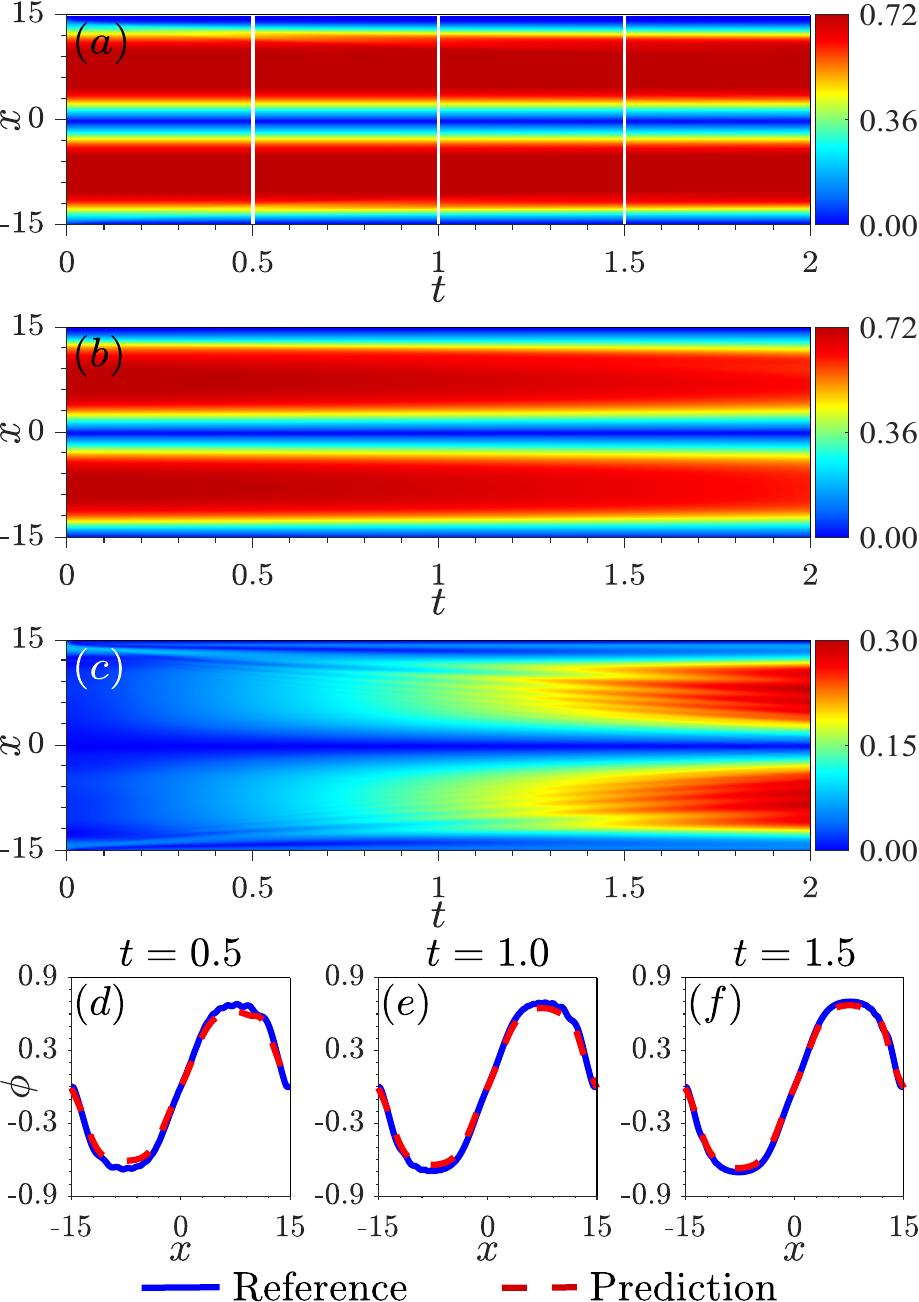}% Here is how to import EPS art
    \caption{The training results of flat-top dipole droplet. (a, b) The reference and learned solutions, respectively. (c) The squared error between the reference and learned solution. Panels (d)-(f) illustrate the comparisons between the reference solution and PINNs results at indicated value of time.}
    \label{fig33}
\end{figure}

By further increasing the number of atoms, the dipole QDs gradually transform into those of the flat-top type. Note that, unlike the flat-top dipole QDs, bound states of two separated identical QDs (``molecules") do not exist in the flat-top regime \cite{mohammedelhadjQuantumDropletMolecules2024}. Accordingly, we have also applied the PINN method to predict flat-top dipole QDs. The evolution of the reference QDs and the predicted ones, for a fixed chemical potential, $\mu =-0.2$, are shown in Figs.~\ref{fig33}(a) and (b), respectively. The absolute error between the exact and predicted solutions, $|\widehat{\phi }(t,x)-\phi (t,x)|$, is plotted in Fig. \ref{fig33}(c), with residuals $\sim $ $10^{-2}$. Figures~\ref{fig33}(d)-(f) present the comparison of the predicted and numerically exact solutions at different times, $t=0.5$, $1$, and $1.5$. The total loss value is minimized to $0.0039\times 10^{-4}$, and the network achieves a relative $\mathbb{L}_{2}$ error of $3.025\times 10^{-2}$ for $z=0$ at time $550.5312$ seconds.

\begin{figure}
    \centering
    \includegraphics[width=\linewidth]{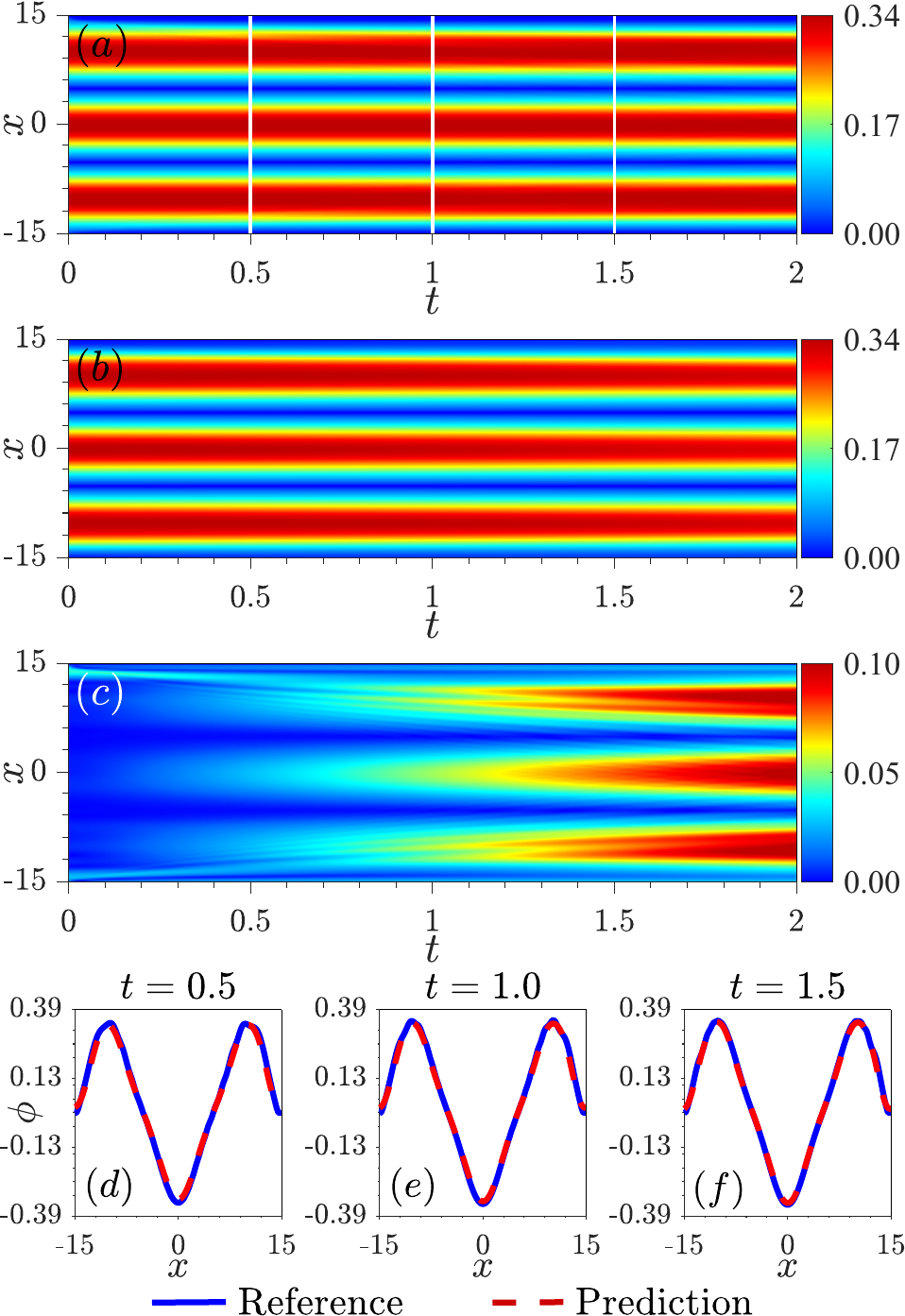}
    % Here is how to import EPS art
    \caption{The PINN results for learning tripole QDs. (a, b) The reference and predicted solutions, respectively. (c) The squared error between the reference and predicted solutions. (d)-(f) The comparison of profiles of the solutions at different times.}
    \label{fig4}
\end{figure}

Tripole QDs feature a rectilinear multipeak structure composed of three fundamental droplets with opposite signs. As a result of the enhanced repulsion between adjacent components, the density distribution in the tripole QDs becomes narrower. The evolution produced by the split-step simulations is shown in Fig.~\ref{fig4}(a), and its counterpart provided by the PINN method is shown in Fig.~\ref{fig4}(b).

The PINNs sampling dataset for the tripole QDs is also obtained using the Newton-conjugate and split-step methods. The spatial and temporal ranges, along with the dataset, are consistent with those used for the monopole QDs, ensuring the uniformity across the entire space-time domain. To approximate the solution $\phi (t,x)$, we employ a neural network with five hidden layers, each consisting of $30$ neurons (like it was done above for the fundamental QDs). For the initial-boundary conditions, we select $N_{0}=400$ from the dataset as sampling points. Additionally, the LHS strategy is utilized to generate $20,000$ collocation points across the entire coordinate domain.

The comparison between the numerically exact tripole QDs solution, $\phi (t,x)$, and the PINN prediction, $\widehat{\phi }(t,x)$, is shown in Figs.~\ref{fig4}(c) in terms of the absolute error, $|\widehat{\phi }(t,x)-{\phi }(t,x)|$. Snapshots of both the numerically exact and PINN-predicted solutions at various times are presented in Figs.~4(d)-(f). After $\sim 20000$ iterations, the total loss function defined by Eq.~\eqref{Loss} is minimized to \textrm{0.42403975}$\times 10^{-3}$, indicating high accuracy. Additionally, the relative $\mathbb{L}_{2}$ error for $z=0$ is found to be \textrm{0.05628608}, achieved at time $703.3468$ seconds. The error between the PINN-predicted solution and its counterpart produced by the split-step method is $\sim 10^{-2}$, demonstrating that the PINN predictions are very close to their counterparts provided by the split-step method. This conclusion suggests that the PINN approach yields highly accurate results with a minimal error, offering a promising alternative to the traditional numerical methods with comparable or even faster performance.

\begin{figure}
    \centering
    \includegraphics[width=\linewidth]{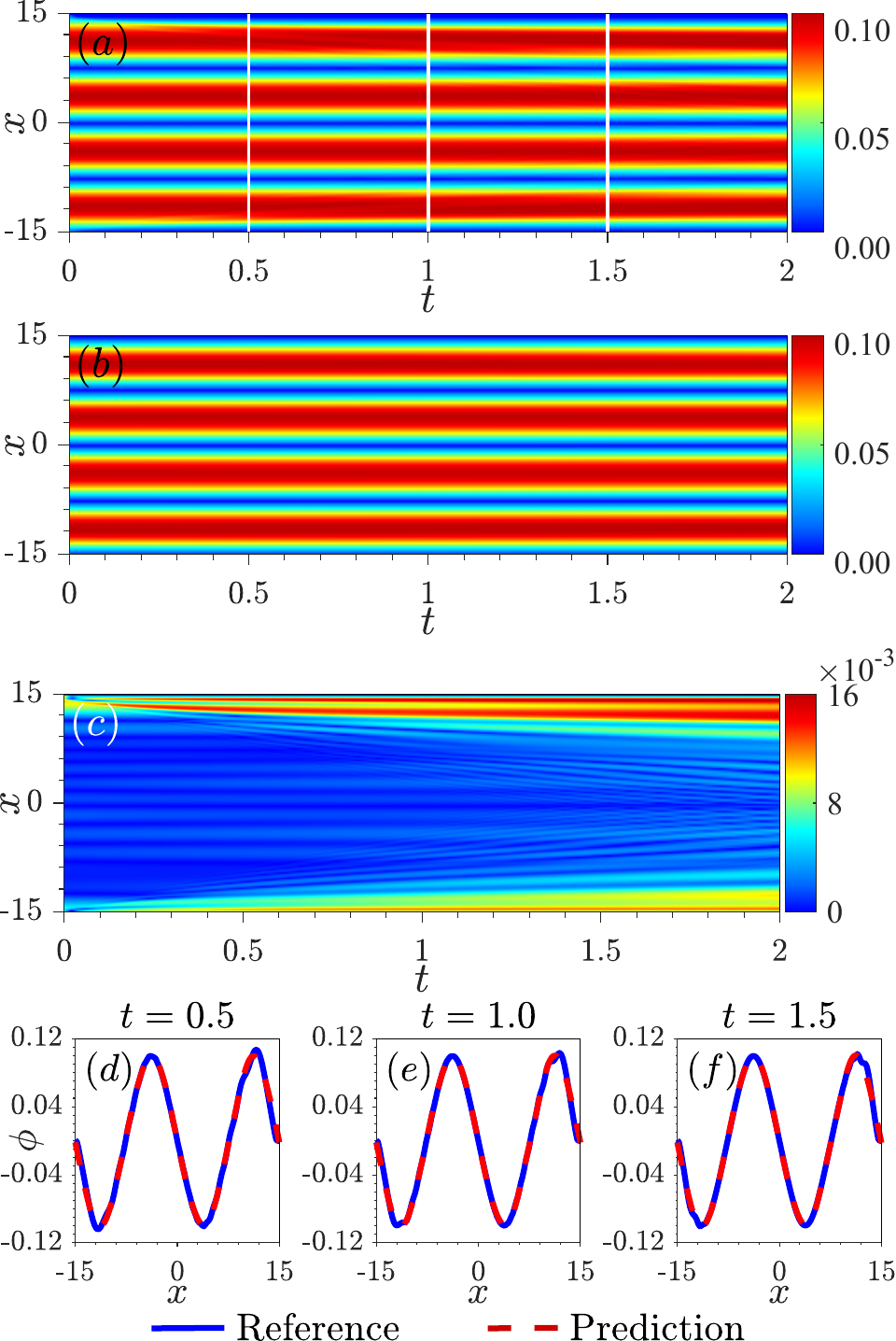}
    % Here is how to import EPS art
    \caption{The PINN results for learning the quadrupole QDs. (a, b) The reference and predicted solutions, respectively. (c) The squared error between the reference and predicted solutions. (d)-(f) The profiles of the solutions at different times.}
    \label{fig5}
\end{figure}

Proceeding to quadrupole QDs, typical density distributions in them are displayed in Fig.~\ref{fig5}, for chemical potential $\mu =1.3$. Within the same solution window as above, we used the five-hidden-layer neural networks with $30$ neurons per hidden layer, combined with the $\mathrm{tanh}$ activation function, to characterize the solution $\phi (t,x)$ by minimizing the loss value. The respective training set consists of $N_{0}=625$ randomly sampled points that form the initial BCs, accounting for approximately $\sim 4.8\%$ of the total data. According to the definition given by Eq.~\eqref{Freim}, the number of collocation points $F_{1}(t,x)$ for PINN is set to be $N_{\mathrm{F}}=20000$. All sampling points are set across the entire space-time domain, using the LHS strategy. For the Adam and multi-step L-BFGS optimization methods, the number of iterations is fixed to be $20000$.

The evolution of quadrupole QDs, as derived by the split-step method, is shown in Fig.~\ref{fig5}(a), with the PINN-predicted counterpart presented in Fig.~\ref{fig5}(b). Owing to the enhanced repulsion between constituents of multipole QDs, the effective width of the individual peaks of quadrupoles significantly decreases. The quadrupole QDs are evenly distributed across the entire window range and are stable.

After $10000$ steps of the Adam optimization, followed by the L-BFGS refinement, the network achieves a relative $\mathbb{L_{2}}$ error of $0.0543$ at time $727.0328$ seconds, with a total loss value of $0.2171\times 10^{-4}$. The absolute error between the PINN prediction and numerically exact solution is found to be $\sim 10^{-3}$, as shown in Fig.~\ref{fig5}(c). Snapshots of the density distribution in the quadrupole QD at times, $t=0.5$, $1.0$, and $1.5$, are depicted in Figs.~\ref{fig5}(d)-(f). The PINN-predicted solutions and their numerical counterparts nearly completely overlap, further confirming that the PINN approach can be used to accurately predict the evolution of the QDs in the binary BEC.

\begin{figure}
    \centering
    \includegraphics[width=\linewidth]{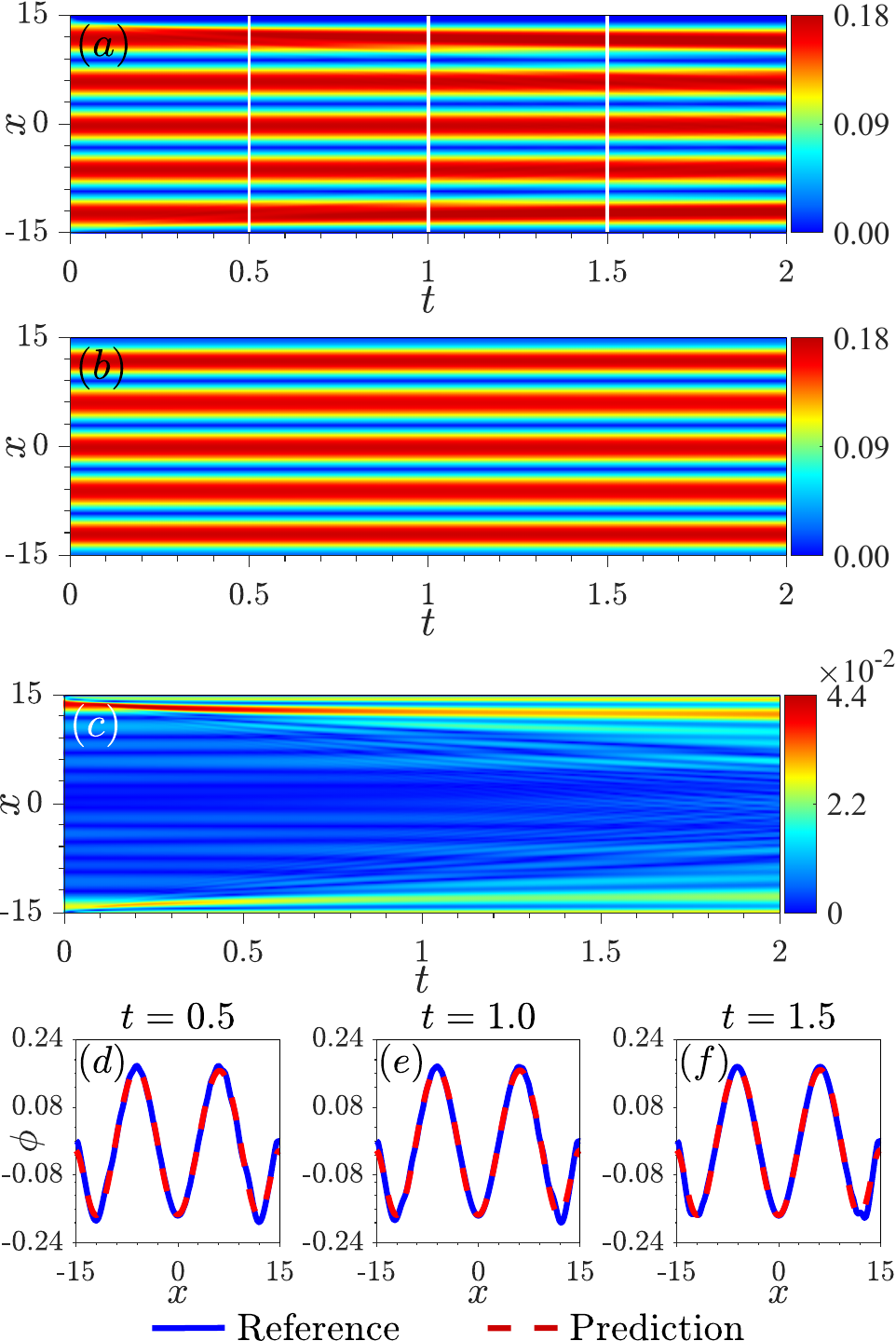}% Here is how to import EPS art
    \caption{The PINN results for learning the pentapole QD. (a, b) The reference and predicted solutions, respectively. (c) The squared error between the reference and predicted solutions. (d)-(f) The profiles of the solutions at different times.}
    \label{fig6}
\end{figure}

Further, the evolution of the pentapole QD, with chemical potential $\mu =1.3$, is depicted in Figs.~\ref{fig6}(a) and (b), as produced by the split-step method and the PINN technique, respectively. The entire space-time domain $(t,x)\in \lbrack 0,2]\times \lbrack -15,+15]$ was divided in $101\times 128$ intervals in both the $t$ and $x$ directions. The loss function, given by Eq.~\ref{Loss}, is minimized by the PINN method, which employs a five-layer neural networks with $10$ neurons per layer. We randomly chose $N_{0}=1600$ points from the initial-boundary conditions, and $N_{\mathrm{F}}=20000$
collocation points were generated using the LHS technique across the space-time domain.

After $20000$ iterations of the Adam optimizer and L-BFGS refinement, the network achieved a relative $\mathbb{L_{2}}$ error of $0.0749$, in approximately $440.0839$ seconds, with the total loss close to $0.2417\times 10^{-3}$. The absolute difference between the predicted and exact solutions for the pentapole QDs is shown in Fig.~\ref{fig6}(c), being $\sim $ $10^{-2}$. Snapshots of the density at various times are provided in Figs.~\ref{fig6}(d)-(f), further confirming the close proximity between the PINN-predicted and numerically exact solutions.

\begin{figure}
    \centering
    \includegraphics[width=\linewidth]{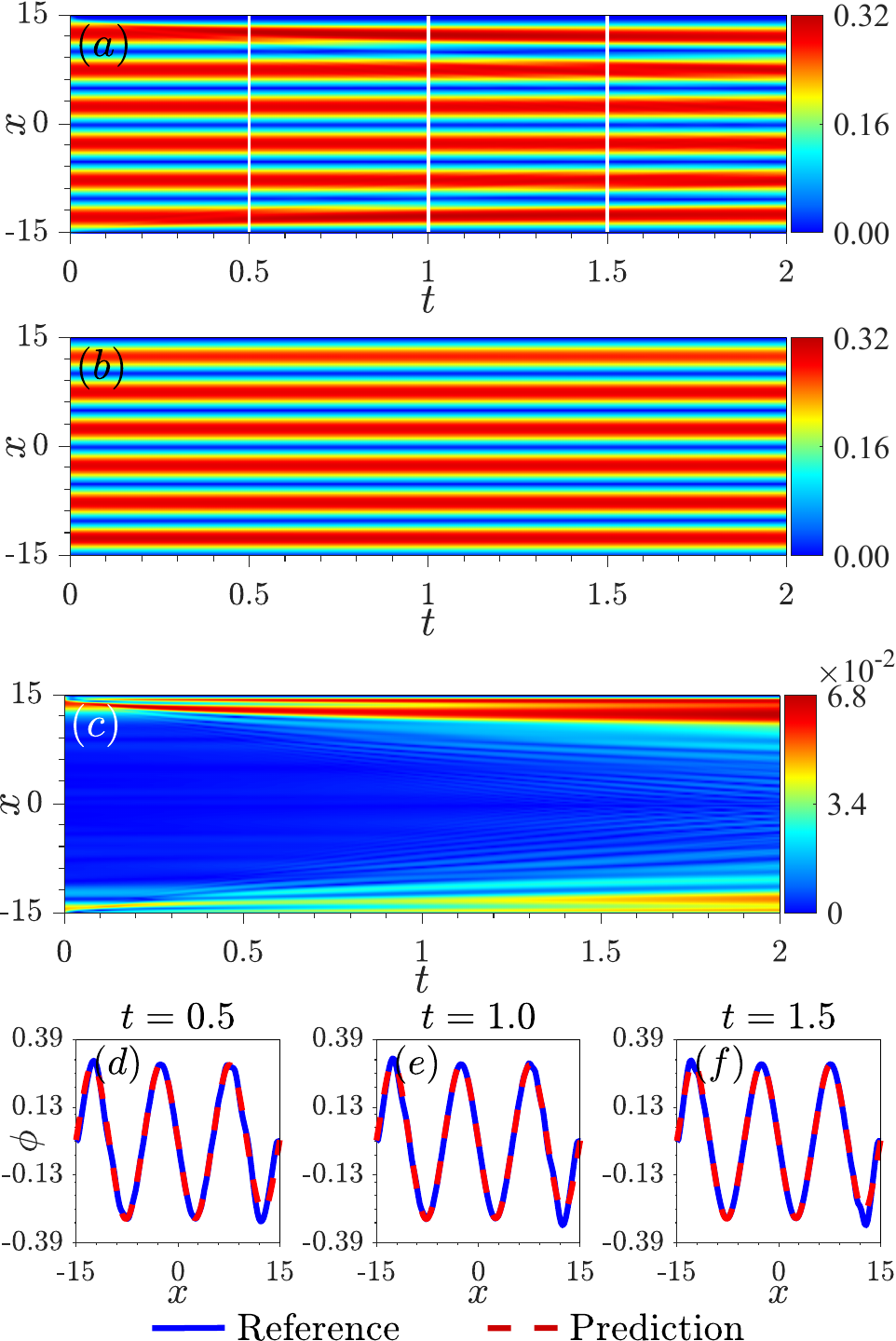}
    % Here is how to import EPS art
    \caption{The PINN\ results for learning the sextupole droplet. (a, b) The reference and predicted solutions, respectively. (c) The squared error between the reference and predicted solutions. (d)-(f) The profiles of the solutions at different times.}
    \label{fig7}
\end{figure}

With an increasing number of droplet constituents (poles), the evolution of the sextupole QDs is shown in Fig.~\ref{fig7}. The PINN sampling dataset is obtained using the Newton-conjugate and the split-step methods, for $\mu =1.6$. The entire space-time domain $(t,x)\in \lbrack 0,2]\times \lbrack -15,15]$ is discretized into $101\times 128$ equal intervals in the $t$ and $x$ directions. For the initial-boundary conditions, $\approx 4.8\%$ of the dataset is selected as sampling points. As in the case of the sextupole, we utilized a five-layer deep neural network, with $20$ neurons per layer. The LHS strategy is employed to generate $N_{\mathrm{F}}=20000$ collocation sampling points across the entire coordinate domain.

The result produced by the split-step method for the sextupole QDs is shown in Fig.~\ref{fig7}(a), and the corresponding PINN-predicted result is shown in Fig.~\ref{fig7}(b). After $20000$ steps of the Adam optimization, followed by the L-BFGS refinement, the network achieves a minimal loss value of $0.3261\times 10^{-3}$. The relative $\mathbb{L_{2}}$ error was $0.0775$, produced by the optimization process taking $588.92$ seconds. The absolute error between the two methods is $\sim 10^{-2}$, as shown in Fig.~\ref{fig7}(c). Figures~\ref{fig7}(d)-(f) further corroborate the near-perfect agreement between the numerically exact and PINN-predicted solutions for the sextupole QDs at different times. From the density maps provided, it is evident that the optimized PINN predictions match the reference solution very closely across the entire space-time domain.

The structure of neural networks is not always the same for different problems. In general, networks with more hidden layers and more neurons per layer tend to perform better. To evaluate the PINN performance in predicting the evolution of QDs under the action of the MF nonlinearity and LHY correction to it, in the framework of the GPE, we explored its predictive capability by varying the number of layers in the neural network. Specifically, we fixed the number of neurons per layer to be $20$ and analyzed the impact of the number of hidden layers on the PINN's prediction ability by comparing various metrics, including the training time, loss value, and $\mathbb{L_{2}}$ error.

The results from multiple predictions are collected in Table~\ref{table:1}. By comparing networks with different numbers of hidden layers, we find that the PINN with $5$ hidden layers achieves the smallest $\mathbb{L_{2}}$ error, while also having a relatively small total loss and training time. For a network with only one hidden layer, after $20000$ iterations of the Adam optimizer, followed by L-BFGS refinement, the relative $\mathbb{L_{2}}$ error is $0.0624$, achieved in approximately $403.0464$ seconds, with a total loss close to $0.4243\times 10^{-6}$. When the number of hidden layers (PINN items, d) exceeds $8$, the increase in the storage space and training time results in only marginal improvements in the model's performance.

\begin{table}
    \centering
        \fontsize{6}{7} \selectfont    %{Font Size}{line spacing}
        \caption{The comparison between the depths of the neural networks.}
            \begin{tabular}{l|cccc}
            \toprule
            \diagbox [width=7em,trim=l] {Data}{Items(d)} & 1 & 2 & 5 & 8 \\ \hline
            Training time & 403.0464 s & 646.2643 s & 690.9437 s &  1356.6818 s \\
            Loss value & 4.243164e-07 & 4.810588e-07 & 2.310483e-07 & 2.013381e-07 \\
            $\mathbb{L_2}$ error (t=0) & 6.244839e-02 & 6.328273e-02 & 3.806830e-02 & 3.999140e-02 \\ 
            \bottomrule
            \end{tabular}
        \label{table:1}
\end{table}

The above results show that the PINN method with fewer layers can still produce relatively accurate solutions, highlighting its significant advantage in predicting the evolution of QDs in the binary BEC. Considering the trade-off between the accuracy and computational complexity, we opt for the $5$-layer network to predict the evolution dynamics of multipole QDs.

%%%%%%%%%%%%%%%%%%%%%%%%%%%%%%%%%%%%%%%%%%%%%%%%%%%%%%%%%%%%%%%%%%%%%%%%%%%%%%%%%%%%

\section{Data-driven parameter discovery}

\label{}

In this section, we address the discovery of data-driven parameters in the $(1+1)$-dimensional GPE using the PINN method. The data-driven parameter inversion is crucial for identifying models in various applications. Traditional numerical methods face certain challenges in discovering model
parameters. A promising approach is to leverage fundamental physics to identify unknown coefficients with physical meaning from auxiliary measurements. In this section, we treat the coefficients of the governing equations as unknowns. The GPE with unknown parameters \eqref{eq:refname4} is expressed as follows:
\begin{equation}
{i}\frac{\partial \phi }{\partial {t}}+\lambda _{1}\frac{\partial ^{2}\phi }{%
\partial x^{2}}+\lambda _{2}|\phi |\phi -|\phi |^{2}\phi =0,  \label{F2}
\end{equation}%
where $\phi (t,x)=u(t,x)+iv(t,x)$ is the normalized complex QD wave function. Parameters $\lambda _{1}$ and $\lambda _{2}$ pertain to the kinetic-energy and LHY-correction terms, respectively.

As is well known and mentioned above, the formation of QDs in binary BEC results from the balance between the MF interactions and LHY correction to it \cite{kartashovMultipoleQuantumDroplets2024a,petrovQuantumMechanicalStabilization2015g}. Therefore, an important QD characteristic arises from different effects of the LHY correction in different dimensions. As said above, in the case of the effectively one-dimensional BEC, the LHY correction corresponds to the effective attraction that compensates for the MF intra-species repulsion, enabling the formation of self-bound QDs even in free space \cite{petrov_ultradilute_2016}. The competition between the MF and LHY nonlinearities leads to shape transformations and stability in QDs. In this context, we use the PINN method to invert the problem and predict the kinetic and nonlinearity coefficients $\lambda _{1}$ and $\lambda _{2}$ in Eq. \eqref{F2}.

Similar to the forward problem, we first define the residual equation for the neural network $\widehat{\phi }(t,x,\lambda _{1},\lambda _{2})$ as:
\begin{equation}
F_{2}(t,x)\equiv i\frac{\partial \phi }{\partial {t}}+\lambda _{1}\frac{%
\partial ^{2}\phi }{\partial x^{2}}+\lambda _{2}|\phi |\phi -|\phi |^{2}\phi
,  \label{Finverse}
\end{equation}%
cf. Eq.~\eqref{F2}. In Eq. \eqref{Finverse}, the residual neural network shares the same unknown parameters as the underlying GPE \eqref{F2}. Next, we introduce complex fields into the residual neural network, leading to the complex-valued PINN formulation, $F_{2}(t,x)=F_{u}(t,x)+iF_{v}(t,x)$,
expressed as
\begin{equation}
\begin{split}
& F_{u}=-\widehat{v}_{z}+\lambda _{1}\frac{\partial ^{2}\widehat{u}}{%
\partial x^{2}}+\lambda _{2}\sqrt{\widehat{u}^{2}+\widehat{v}^{2}}\widehat{%
u}-(\widehat{u}^{2}+\widehat{v}^{2})\widehat{u}, \\
& F_{v}=-\widehat{u}_{z}+\lambda _{1}\frac{\partial ^{2}\widehat{v}}{%
\partial x^{2}}+\lambda _{2}\sqrt{\widehat{u}^{2}+\widehat{v}^{2}}\widehat{%
v}-(\widehat{u}^{2}+\widehat{v}^{2})\widehat{v}.
\end{split}
\label{eq:refname13}
\end{equation}%
To train the PINN neural network $\widehat{\phi }(t,x,\lambda _{1},\lambda _{2})$ and estimate the unknown parameters $\lambda _{1}$ and $\lambda _{2}$, we define the following loss function:
\begin{equation} 
\begin{split}
\mathrm{MSE}& =\frac{1}{N}\sum_{j=1}^{N}[|\widehat{u}%
(t^{j},x^{j})-u^{j}|^{2}+|\widehat{v}(t^{j},x^{j})-v^{j}|^{2} \\
& \quad +|F_{u}(t^{j},x^{j})|^{2}+|F_{v}(t^{j},x^{j})|^{2}],
\end{split}
\label{eq:refname14}
\end{equation}%
where $\{t^{j},x^{j},\phi ^{j}\}_{j=1}^{N}$ represents a set of randomly sampled data points generated by the split-step solution. The neural network structure for the inverse problem is essentially similar to that for the forward problem, with the key distinction lying in the choice of the loss function.

Building on the forward problem investigated above, we now utilize sextupole QDs as the sampling dataset. The sampling points are generated using the numerically exact sextupole solutions of Eq.~\eqref{F2} with parameters $\lambda _{1}=0.5$ and $\lambda _{1}=1$. To perform the parameter inversion, we randomly select $N=625$ data points across the entire space-time domain $(t,x)\in \lbrack 0,2]\times \lbrack -15,15]$, which constitute approximately $4.8\%$ of the total data. Initially, the unknown parameters are set to $\lambda _{1}=\lambda _{2}=0$ (different from the above-mentioned values $\lambda _{1}=0.5$, $\lambda _{1}=1$, which were used to generate the sampling points). The neural network architecture consists of five hidden layers, each containing $30$ neurons. The network is trained using the Adam optimizer for $20000$ steps, followed by $20000$ steps of the L-BFGS optimization algorithm. In the noisy scenario, the same network structure and sampling points are employed for training, ensuring consistency in the evaluation.

The training results for the unknown parameters $\lambda _{1}$ and $\lambda _{2}$ are summarized in Table \ref{table:2}, covering various scenarios and their respective training errors. In the absence of noise, the estimated values of $\lambda _{1}$ and $\lambda _{2}$ are learned as $0.50660$ and $0.60445$, respectively, yielding relative errors of $1.32015\%$ and $11.63909\%$. When $1\%$ random noise is introduced into the training data, the parameters are identified as $0.44180$ for $\lambda _{1}$ and $0.95157$ for $\lambda _{2}$, with relative errors increasing to $11.63909\%$ and $4.84328\%$, respectively. Despite the presence of noise, the results indicate that the unknown parameters can still be accurately recovered with remarkable precision. This highlights the ability of the PINN framework to reliably predict parameters of the physical model while utilizing a relatively small subset of the data, constituting only $\approx 4.8\%$ of the total points. These findings emphasize the robustness and efficacy of incorporating physical principles into neural networks, demonstrating the potential of this approach for solving complex inverse problems with noisy data.

\begin{table}[htbp!]
    \centering
     \fontsize{6}{7} \selectfont    %{Font Size}{line spacing}
     \caption{The data-driven parameter discovery for $\left\{ \protect\lambda _{1},\protect\lambda _{2}\right\}$ (see Eq. \protect\eqref{F2}) and the corresponding errors.}
         \begin{tabular}{l|cccc}
             \toprule
         \diagbox [width=13em,trim=l] {Data}{Value and error} & $\lambda_{1}$ & $\lambda_{2}$ & error of $\lambda_{1}$ & error of $\lambda_{2}$ \\ \hline
         Correct data & 0.5 & 1 & 0 &  0 \\
         Clean training data & 0.50660 & 1.00604 & $1.32015\%$ & $0.60445\%$ \\
         Training data with $1\%$ noise & 0.44180 & 0.95157 & $11.63909\%$ & $4.84328\%$ \\
     \bottomrule 
     \end{tabular}
     \label{table:2}
 \end{table}

Note that the addition of noise to the training data leads to a significant amplification of parameter-estimation errors in PINNs. This phenomenon primarily stems from two factors: the interference of noise with the optimization process, and the potential failure of the physics-based constraints to effectively regularize the corrupted data. In essence, the noise disrupts the precise learning of the underlying physical laws embedded in the loss function. Consequently, the model cannot correct the deviations induced by the noise, resulting in inaccurate predictions and increased errors. To mitigate this issue, the increase of the volume of the training data can enhance the model's robustness by providing a better approximation of the true data distribution.
%%%%%%%%%%%%%%%%%%%%%%%%%%%%%%%%%%%%%%%%%%%%%%%%%%%%%%%%%%%%%%%%%%%%%%%%%%%%%%%%%

\section{Conclusion}\label{}

This study demonstrates the powerful capability of PINNs in analyzing the evolution of QDs (quantum droplets) and the parameter discovery for them in binary BECs. Integrating deep learning with the prior physical knowledge, PINNs provide an effective framework for studying the complex mean-field GPEs for macroscopic quantum systems. The results highlight the accuracy of PINNs in predicting the structural features, such as distinct peak profiles and the stable evolution of multipole QDs, as well as their evolution in time. Through systematic comparison of different neural network architectures, we show that even networks with fewer layers can reliably predict the evolution of QDs, achieving low values of the $\mathbb{L_{2}}$ error and minimal loss values. Additionally, the PINN robustness is validated by the solution of the parameter-discovery problem, demonstrating its efficiency in identifying unknown coefficients of the underlying GPE (Gross-Pitaevskii equation), even when working with to noisy data. With the added complexity of multipole QDs exhibiting multipeak structures, PINN accurately predicts their stable evolution and reveals effects of the LHY correction.

These findings underscore the versatility and robustness of employing PINNs for solving complex forward problems, modeling the QD dynamics and parameter identification in binary BECs, including those under the action of noisy environments.

%%%%%%%%%%%%%%%%%%%%%%%%%%%%%%%%%%%%%%%%%%%%%%%%%%%%%%%%
\medskip

%\begin{backmatter}

% \section*{CRediT authorship contribution statement}
% \textbf{Dongshuai Liu}: Numerical calculations, Writing - original draft. 
% \textbf{Wen Zhang}: Numerical calculations, Writing codes. 
% \textbf{Boris A. Malomed}: Writing - review and editing.

\section*{Funding}

This work was partially supported by the Doctoral Foundation of Fuyang Normal University (No. 2025KYQD0080, 2024KYQD0094), the Natural Science Foundation of Fuyang Normal University (No. 2025FSKJ2, 2025FSKJ09) and Excellent Research and Innovation Team of Functional Materials and Devices for Informatics of Anhui Higher Education Institute (No. 2024AH010024).

\section*{Declaration of competing interest}

The authors declare no potential conflict of interest.

\section*{Data availability}

Research data will be made available upon a reasonable request.

\bibliographystyle{elsarticle-num-names} 
\bibliography{reference}

\end{document}